\documentclass[english,aps,prb,showpacs,superscriptaddress,floats,amsmath,amssymb,floatfix,nobalancelastpage,twocolumn,groupedaddress]{revtex4-2}
\usepackage[T1]{fontenc}
\usepackage[latin9]{inputenc}
\usepackage{units}
\usepackage{dsfont}
\usepackage{amsmath}
\usepackage{amssymb}
\usepackage{graphicx}

\makeatletter

\newcommand*\LyXThinSpace{\,\hspace{0pt}}

\usepackage{accents}

\makeatother

\usepackage{babel}
\graphicspath{{./}{./fig/}{./figs/}{./psfigs/}{./Figures/}}
\begin{document}
\title{Two-dimensional nonlinear dynamical response of the\\
magnetoelectrically driven dimerized spin-$1{/}2$ chain}
\author{Wolfram Brenig}
\email{w.brenig@tu-braunschweig.de}

\affiliation{Institute for Theoretical Physics, Technical University Braunschweig,
D-38106 Braunschweig, Germany}
\begin{abstract}
A study of the dynamical two-dimensional (2D) nonlinear response of
the dimerized spin-$1{/}2$ chain to external electric fields is
presented. The coupling of the spin system to those fields is set to arise
from the inverse Dzyaloshinskii-Moriya interaction. In the XY-limit,
we provide analytical expressions for the second-order nonlinear dynamical
response function. Apart from multi spinon continua, this response
displays a strong antidiagonal, i.e. galvanoelectric, feature in the
2D frequency plane. This allows to read off scattering rates of the
fractional spinon excitations. For the XXZ-case, we focus on the interaction-driven
renormalization of the light-matter coupling by considering vertex
corrections which are induced by the zz-exchange. We show this renormalization
to modify the spinon joint density of states significantly, potentially
allowing for the formation of in-gap bound states. As a result, the
vertex corrected light-matter coupling can induce a dramatic spectral
weight transfer to lower energies for the dynamical response function
within the 2D frequency plane.
\end{abstract}
\maketitle

\section{Introduction\label{sec:Introduction}}

Probing magnetism by electric fields, i.e., the magnetoelectric effect
(MEE), has a long standing history \citep{Dzyaloshinskii1960} and
remains of great importance until today \citep{Fiebig2005,Tokura2014,Dong2015,Fiebig2016,Dong2016}.
By now, the MEE is central to a multitude of applications \citep{Liang2021}
which are part of a broad road-map of materials and future devices
\citep{Liang2021b}. Arguably, the main interest in the MEE rests
on the ability to manipulate the static magnetization via static electric
fields.

Beyond purely static phenomena, the dynamics of the MEE is also of
great interest. This is particularly evident in the context of the
so-called electromagnons, which are hybrid-modes of magnons and photons
or polar-phonons \citep{Baryakhtar1970,Smolenskii1982}. Primarily
they have been analyzed in magnetoelectrics with helical forms of
magnetic correlations and order \citep{Pimenov2006,Sushkov2007,Shuvaev2010,Mochizuki2010,Takahashi2012,Toth2016}.
Additional kinds of coupled magnetoelectric excitations can occur in
the presence of more complex magnetic order parameters \citep{Fishman2017,Kocsis2019}.
Coupling to magnons by the MEE can also be seen directly in the optical
conductivity \citep{Sushkov2007,Fang2008,Sushkov2008,Ueda2022}, which
allows for terahertz spectroscopy of magnetic excitations without
the need for circularly polarized fields.

MEE spectroscopy of magnons by optical conductivity is a linear-response
probe. Beyond that, nonlinear optical response of ferroelectrics and
multiferroics \citep{Azimi2016,Zhang2021} is considered as a promising
direction which deserves further pursuit. Among the various nonlinear
spectroscopies \citep{Mukamel1999,Butcher1990}, two-dimensional (2D)
coherent nonlinear optical spectroscopy (2DCS) \citep{Mukamel2000,Brixner2004,Cho2008}
is of particular interest, since by Fourier transforming the nonlinear
response resulting from particular sequences of ultrashort laser pulses
\citep{Mukamel2000,Brixner2004,Cho2008}, spectral information can
be obtained in more than one frequency dimension. In molecules, low-dimensional
semiconductors, and nanomaterials, this has lead to a wealth of information
about structure, electronic and vibrational excitations, dynamics,
relaxation, and dephasing \citep{Cho2008,Huse2005,Cowan2005,Kuehn2009,Kuehn2011}.

Recently many-body phases of magnetic and correlated electronic systems
have come under scrutiny of 2DCS, studying magnons \citep{Lu2017,Kaib2024},
solitons \citep{Wan2019}, Majorana fermions and visons \citep{Choi2020,Negahdari2023,Qiang2023,Brenig2024},
spinons \citep{Li2021,Sim2023L,Gao2023,Hart2023,Sim2023,ZLLi2023,Potts2023,Watanabe2024},
fractons \citep{Nandkishore2021}, marginal Fermi liquids \citep{Mahmood2021},
and the Anderson impurity model \citep{Kappl2023}.
This research is accompanied by significant progress in development
and application of methods for calculating nonlinear response functions
(NRF)s.
Among the various systems considered, 1D quantum magnets, i.e., non-/integrable
spin chains \citep{Wan2019,Sim2023L,Gao2023,Sim2023,Watanabe2024}
and Tomonaga-Luttinger liquids \citep{Li2021,ZLLi2023}, are of particular
interest because fractionalization of spin into
spinons is a hallmark of such magnets. Because of that, linear response
probes of spin excitations in these systems lead to ubiquitous two-particle
continua, notoriously prohibiting access to single-spinon states.
Here, 2DCS has been suggested for deconvoluting
such fractionalization continua \citep{Wan2019}.

Remarkably, the main body of analysis of 2DCS in low-dimensional quantum
magnets reported up to now has focused on the direct coupling of
the magnetic field component of the light to the spin system via the
Zeeman energy. 2DCS from coupling to the electric field has been advocated
only very recently in the context of the Kitaev model \citep{Brenig2024},
for the noninteracting dimerized twisted Kitaev chain \citep{Srivastava2025},
and for the planar $J_{1}$-$J_{3}$ antiferromagnet in the spiral
phase \citep{Brenig2025}. In this work, we will proceed further into
this direction and consider electric-field mediated 2DCS on the
dimerized spin-$1{/}2$ chain, studying separately the XY- and XXZ-case.
For the former we will show that for free spinons, the structure
of the 2D NRF allows to disentangle elementary spinon spectral properties
from the multi-spinon continua which are inevitable because of fractionalization.
For the latter we will make a first step to include spinon interactions
by investigating vertex corrections which generate an effective light-matter
coupling. We find that collective spinon bound states arising from
that may drastically modify the 2D NRF.

The paper is organized as
follows. In Sec. \ref{sec:xxz+knb} we detail the model and its treatment
using Jordan-Wigner fermions. In Sec. \ref{sec:2DCS-Response-Fun}
the evaluation of the 2D NRF in the XY-limit is described, followed
by a discussion of its properties in Sec. \ref{subsec:xydisc}. In
Sec. \ref{sec:renlmc} we explain the vertex corrections for the XXZ-case,
detail their RPA derivation in Sec. \ref{subsec:RPA}, and discuss the renormalized
2D NRF in \ref{subsec:xxzdisc}. In Sec. \ref{sec:sum} we summarize.

\section{The dimerized xxz-chain with KNB-coupling to electric fields\label{sec:xxz+knb}}

The mechanisms driving the MEE in Mott insulators rest on three microscopic
processes at large \citep{Tokura2014}. The first is the spin-dependent
hybridization \citep{Jia2007} which is strictly zero for spin-$\nicefrac{1}{2}$
system, i.e. for our study. Second, exchange-striction coupling can
occur \citep{Katsura2009}, requiring however some sort of broken
inversion symmetry of the system which we discard for this work.
Finally, the spin-current or inverse Dzyaloshinskii--Moriya (DM) interaction
introduced by Katsura, Nagaosa, and Balatsky (KNB) \citep{Katsura2005}
is an option. This coupling has been under intense scrutiny in many
quasi-1D spin models in the context of electric-field control of quantum
phases, in particular since it can be treated more or less exactly
in several 1D situations \citep{Brockmann2013,You2014,Menchyshyn2015,Thakur2018,Baran2018,Yi2019,Strecka2020,Richter2022}.

In this work we focus on a dimerized spin-$1{/}2$ chain in the presence
of KNB coupling to electric fields
\begin{align}
H/J= & \sum_{l}[(1+(-)^{l}\delta)(S_{l}^{x}S_{l+1}^{x}+S_{l}^{y}S_{l+1}^{y}+\Delta S_{l}^{z}S_{l+1}^{z})-\nonumber \\
 & \hphantom{\sum_{l}[}g_{l}\boldsymbol{E}\cdot\boldsymbol{e}_{l,l+1}\times(\boldsymbol{S}_{l}\times\boldsymbol{S}_{l+1})]\,.\label{eq:model}
\end{align}
The chain is directed along the $x$-axis, and the addend $\propto g_{l}$
encodes the KNB coupling with a bond-dependent strength $g_{l}$.
It displays the cross-products characteristic of DM interactions,
with $\boldsymbol{e}_{l,l+1}$ being a unit vector along $x$. The
electric field $\boldsymbol{E}=\boldsymbol{E}_{dc}+\boldsymbol{E}_{ac}(t)$
is homogeneous in space and is decomposed into a dc contribution $\boldsymbol{E}_{dc}$
and a part $\boldsymbol{E}_{ac}(t)$ which time-averages to zero.
Energies are normalized to an exchange constant $J$, which is set
to unity henceforth $J=1$. $\delta$ sets the size of the exchange
dimerization and we assume the KNB coupling to be dimerized likewise,
with $g_{l}=1+(-)^{l}\gamma$. Because $g_{l}=1$ for $\gamma=0$
and in view of the combination $g_{l}\boldsymbol{E}$ in Eq. (\ref{eq:model}),
the definition of $\boldsymbol{E}$ is assumed to absorb a dipole moment
type coupling constant as a free parameter. We have
\begin{equation}
\boldsymbol{e}_{l,l+1}\times(\boldsymbol{S}_{l}\times\boldsymbol{S}_{l+1})=\begin{bmatrix}0\\
S_{l}^{y}S_{l+1}^{x}-S_{l}^{x}S_{l+1}^{y}\\
S_{l}^{z}S_{l+1}^{x}-S_{l}^{x}S_{l+1}^{z}
\end{bmatrix}\,.\label{eq:knbs}
\end{equation}
By symmetry, applying the electric field perpendicular to the chain
does not discriminate between $\boldsymbol{E}\parallel y$ or $z$.
Mathematically, however, since we will use the Jordan-Wigner transformation
to map the spins to fermions, the 2nd and 3rd row of Eq. (\ref{eq:knbs})
are of very different complexity. Only the former will not contain
fermion strings. Therefore, we confine ourselves to
$\boldsymbol{E}=E\boldsymbol{e}_{y}=(E_{dc}+E_{ac}(t))\boldsymbol{e}_{y}$ yielding
\begin{align}
H= & H^{xxz}-E\,P\label{eq:modabr}\\
P= & -\sum_{l}(1+(-)^{l}\gamma)(S_{l}^{x}S_{l+1}^{y}-S_{l}^{y}S_{l+1}^{x})\,,
\end{align}
where $H^{xxz}$ refers to the right hand side on the first line of
Eq. (\ref{eq:model}) and $P$ labels the $x$-component of the polarization.
Finally, we will absorb the dc-component of the electric field into
the $xxz$-Hamiltonian henceforth, defining $\tilde{H}^{xxz}=H^{xxz}-E_{dc}\,P$.

We now proceed to the Jordan-Wigner representation of the model in
terms of spinless fermions $a_{l}^{(\dagger)}$ at site $l$, and
$S_{l}^{+}=\prod_{l'}^{l-1}(1-2n_{l'})a_{l}^{\dagger}$, as well as
$S_{l}^{z}=n_{l}-1{/}2$, and $n_{l}=a_{l}^{\dagger}a_{l}^{\phantom{\dagger}}$.
Because of the dimerization, the unit cell comprises two sites, implying
a Brillouin zone (BZ) of $k\in[\allowbreak-\pi/2,\allowbreak\pi/2[$
and two types of fermions $a_{1k}^{(\dagger)}$, $a_{2k}^{(\dagger)}$
in momentum space. We obtain
\begin{align}
\tilde{H}^{xxz}= & \tilde{H}^{xy}+H^{zz}=\sum_{k}\begin{bmatrix}a_{1k}^{\dagger} & a_{2k}^{\dagger}\end{bmatrix}\mathds{h}_{k}\begin{bmatrix}a_{1k}^{\phantom{\dagger}}\\
a_{2k}^{\phantom{\dagger}}
\end{bmatrix}+\nonumber \\
 & \sum_{l}\Delta(1+(-)^{l}\delta)(n_{l}-\tfrac{1}{2})(n_{l+1}-\tfrac{1}{2})\label{eq:jwh0}\\
\mathds{h}_{k}= & \begin{bmatrix}0 & t_{k}\\
t_{k}^{\star} & 0
\end{bmatrix}
\end{align}
with $t_{k}=(\cos(k)-E_{dc}\sin(k))+i(\delta\sin(k)+E_{dc}\gamma\cos(k))\allowbreak=A_{k}+iB_{k}$.
The $xy$-part can be diagonalized by a unitary transformation $\mathbb{U}_{k}$
onto fermions $d_{1k}^{(\dagger)}$, $d_{2k}^{(\dagger)}$, with
\begin{equation}
\begin{bmatrix}a_{1k}^{\phantom{\dagger}}\\
a_{2k}^{\phantom{\dagger}}
\end{bmatrix}=\mathbb{U}_{k}\begin{bmatrix}d_{1k}^{\phantom{\dagger}}\\
d_{2k}^{\phantom{\dagger}}
\end{bmatrix},\hphantom{aa}\mathbb{U}_{k}=\tfrac{1}{\sqrt{2}}\begin{bmatrix}\frac{t_{k}}{\epsilon_{k}} & -\frac{t_{k}}{\epsilon_{k}}\\
1 & 1
\end{bmatrix}\label{eq:U}
\end{equation}
and
\begin{equation}
\tilde{H}^{xy}=\sum_{k,\mu=1,2}\epsilon_{\mu,k}d_{\mu k}^{\dagger}d_{\mu k}^{\phantom{\dagger}}\,,\label{eq:xydig}
\end{equation}
and a quasiparticle dispersion $\epsilon_{1,k}=-\epsilon_{2,k}=\epsilon_{k}$
\begin{align}
\epsilon_{k} & =\sqrt{|t_{k}|^{2}}=[(1+E_{dc}^{2}\gamma^{2})\cos(k)^{2}+(\delta^{2}+E_{dc}^{2})\sin(k)^{2}-\nonumber \\
 & \hphantom{aaaaaaaaaaa}E_{dc}(1-\delta\gamma)\sin(2k)]^{1/2}\,.\label{eq:disp}
\end{align}
We note that the last term in the expression for $\epsilon_{k}$
on the second line of Eq. (\ref{eq:disp}), which is proportional
to the dc electric field, is a direct consequence of the broken inversion
symmetry for $E_{dc}\neq0$ and implies $\epsilon_{k}\neq\epsilon_{-k}$
in that case.

Next, we express the KNB dipole moment in terms of the diagonal fermions.
We obtain
\begin{align}
P= & \sum_{k}\begin{bmatrix}d_{1k}^{\dagger} & d_{2k}^{\dagger}\end{bmatrix}\begin{bmatrix}p_{11k} & p_{12k}\\
p_{12k}^{\star} & -p_{11k}
\end{bmatrix}\begin{bmatrix}d_{1k}^{\phantom{\dagger}}\\
d_{2k}^{\phantom{\dagger}}
\end{bmatrix},\label{eq:mc}\\
p_{11k}= & \tfrac{1}{2\epsilon_{k}}[(1-\gamma\delta)\sin(2k)-\nonumber \\
 & \hphantom{\tfrac{1}{2\epsilon_{k}}[}2E_{dc}(\gamma^{2}\cos(k)^{2}+\sin(k)^{2})]\label{eq:p11}\\
p_{12k}= & -\tfrac{i}{\epsilon_{k}}(\delta\sin(k)^{2}+\gamma\cos(k)^{2})\,.\label{eq:p12}
\end{align}
We note that for $E_{dc}=\delta=\gamma=0$, the quasiparticle dispersion
simplifies to $\epsilon_{k}=\cos(k)$
for $k\in$ BZ and $P$ turns diagonal with $p_{11k}=\sin(k)$.

\section{2D response function in the XY-limit\label{sec:2DCS-Response-Fun}}

In this section we will consider the NRFs in the noninteracting XY-limit,
i.e., for $\Delta=0$. The lowest order 2D response to $E_{ac}(t)$
of the polarization $\langle\Delta P\rangle(t)$ can occur at $O(N=2)$.
It is obtained from the Fourier transform into the 2D frequency plane
of the retarded response function $\tilde{\chi}_{2}(t,t_{1},t_{2})=i^{2}\Theta(t-t_{1})\Theta(t_{1}-t_{2})\allowbreak\langle[[P(t),\allowbreak P(t_{1})],P(t_{2})]\rangle$
\citep{Butcher1990}. The $N$-fold time integrations in the perturbative
expansion of $\langle\Delta P\rangle(t)$
are totally symmetric with respect to permutations of the 
time arguments, which is dubbed intrinsic permutation symmetry \citep{Butcher1990}. Therefore, any $N$-th order contribution to $\langle\Delta P\rangle(t)$
requires an evaluation only of
the fully symmetrized response function
$\chi_{N}(t,t_{1},\allowbreak\dots,t_{n})=\allowbreak\sum_{M}\tilde{\chi}_{N}(t,t_{M(1)},\allowbreak\dots,t_{M(N)})/N!$,
where $M$ labels all permutations. The Fourier transform
of $\chi_{N}(t,t_{1},\allowbreak\dots,t_{n})$ can be obtained from
analytic continuation to the real axis of the Matsubara frequency
transform of the fully connected contractions of the imaginary time
propagator $\chi_{n}(\tau_{n},\dots\tau_{1})=\langle T_{\tau}(\allowbreak P(\tau_{n})\allowbreak\dots P(\tau_{1})P)\rangle$
\citep{Evans1966,Rostami2021}.

\begin{figure}[tb]
\centering{}\includegraphics[width=0.7\columnwidth]{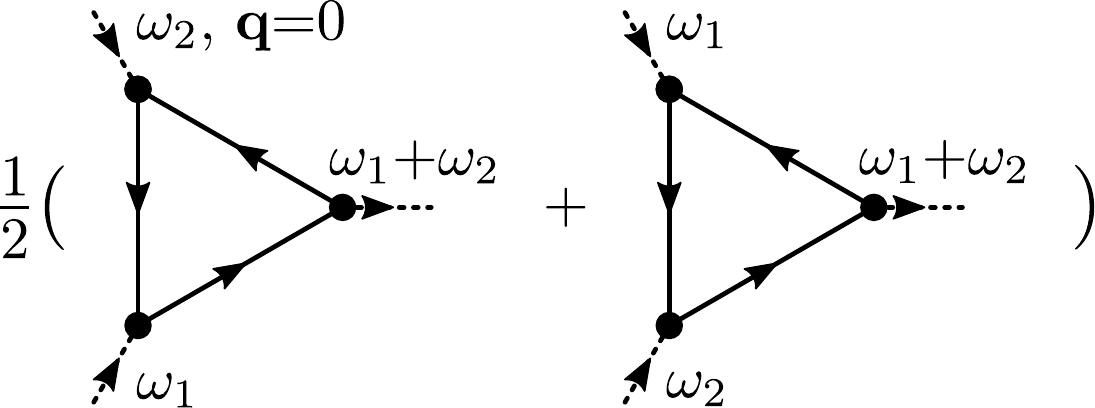}\caption{
Diagrams for the 2D NRF $\chi_{2}(\omega_{1},\omega_{2})$
at order $O(2)$ in $E_{ac}(t)$. The solid lines carry a band index
$\mu=1,2$, the dots refer to the $2\times2$ polarization operator
matrices $p_{\mu\nu}$. I.e., the diagrams correspond to 2$\times$8
expressions.\label{fig:diags}}
\end{figure}

In the XY-limit, i.e., for noninteracting spinons at $\Delta=0$,
and for $N=2$, the two diagrams of Fig. \ref{fig:diags} correspond
to $\chi_{2}$. Evaluating these diagrams is straightforward. After
some algebra, we obtain
\begin{align}
\chi_{2}(\omega_{1},\omega_{2})= & \sum_{k}\left[\frac{8(1-2f_{k})\,p_{11}(k)\,|p_{12}(k)|^{2}\epsilon_{k}^{2}}{(z_{1}^{2}-4\epsilon_{k}^{2})(z_{2}^{2}-4\epsilon_{k}^{2})}\times\right.\nonumber \\
 & \hphantom{\sum aa_{k}}\left.\frac{(z_{1}^{2}+z_{1}z_{2}+z_{2}^{2}-12\epsilon_{k}^{2})}{((z_{1}+z_{2})^{2}-4\epsilon_{k}^{2})}\right]\,,\label{eq:chi2}
\end{align}
with the Fermi function $f_{\mathbf{k}}=1/(e^{\epsilon_{\mathbf{k}}/T}+1)$
and $z_{1,2}=\omega_{1,2}+i\eta$ are complex frequencies, close to
the real axis with $\omega_{1,2}\in\Re$ and $0<\eta\ll$1.

We note that because Eq. (\ref{eq:chi2}) is a particular three-point
Green's function for free fermions, it is no surprise that, formally,
$\chi_{2}(\omega_{1},\omega_{2})$ is fully consistent with results
which have been obtained for 2D NRFs in the Kitaev model \citep{Brenig2024}
- although the latter is a two-dimensional spin model with fermionic
quasiparticles of a different dispersion and dipole matrix elements
of different physical origin and of different momentum dependence.

\subsection{Discussion\label{subsec:xydisc}}

The 2D NRF, Eq. (\ref{eq:chi2}), has four noteworthy properties.
First, and by intrinsic permutation symmetry, it is symmetric under
interchange $\omega_{1}\leftrightarrow\omega_{2}$.

Second, and as $\chi_{2}$ comprises an odd number of polarization operators, it must
vanish identically, if the Hamiltonian, i.e., the density matrix, commutes with
inversion. In our case, the dc electric field breaks inversion symmetry. In fact, if
we set $E_{dc}=0$, the dispersion satisfies $\epsilon_{k}=\epsilon_{-k}$, as well as
the dipole matrix elements $p_{11k}=-p_{11-k}$ and $p_{12k}=p_{11-k}$.  Therefore, if
$E_{dc}=0$, Eq. (\ref{eq:chi2}) produces $\chi_{2}(\omega_{1},\omega_{2})=0$ for any
$\omega_{1,2}$. Using electric dc fields to force a nonlinear response at $O(2)$,
despite inversion symmetry of the original system, is a well known practice. See,
e.g., the field-induced second harmonic generation in semiconductors
\citep{Aktsipetrov1996} or graphene \citep{Bykov2012}.

Third, and also related to the inversion symmetry broken by $E_{dc}$,
we must have that $\epsilon_{-k}(E_{dc})=\epsilon_{k}(-E_{dc})$ and
$-p_{11\,-k}(E_{dc})=p_{11\,k}(-E_{dc})$, where we have made the
dependence of $\epsilon_{k}$ and $p_{\mu\nu k}$ on $E_{dc}$ explicit.
This can also be checked directly from Eqs. (\ref{eq:disp},\ref{eq:p11}).
Therefore, $\chi_{2}(\omega_{1},\omega_{2},E_{dc})=-\chi_{2}(\omega_{1},\omega_{2},-E_{dc})$,
i.e., the response flips signs if the direction of the dc field is
reversed. This meets physical expectations.

Finally, being a three-point Green's function, $\chi_{2}(\omega_{1},\omega_{2})$
allows for a particularly singular line in the $(\omega_{1},\omega_{2})$-plane,
namely the so-called rectification or galvanoelectric (GEE) line,
along which $\omega\equiv\omega_{1}=-\omega_{2}$, rendering the response
completely dc, i.e., $\omega_{1}+\omega_{2}=0$. Along this line and
asymptotically for $\eta\ll1$, we find
\begin{align}
\lefteqn{\chi_{2}(\omega{,}{-}\omega)\simeq\frac{\pi}{\eta}\sum_{k}\left[(1{-}2f_{k})\,p_{11k}\times\right.}\nonumber \\
 & \hphantom{aaaaaaaaaa}\left.|p_{12k}|^{2}(\delta(\omega+2\epsilon_{k})+\delta(\omega-2\epsilon_{k}))\right].\label{eq:c2gee2}
\end{align}
This response function is purely real and comprises two parts: (a)
The terms summed over $k$ which encodes a form-factor weighted interband
density of states. In the limit $\eta\rightarrow0^{+}$, this is a
non-singular, smooth function, except for potential van Hove singularities.
(b) A prefactor $\propto1/\eta$, which renders the GEE response globally
singular for $\eta\rightarrow0^{+}$. The latter behavior has also
been noted for other systems \citep{Parker2019,Sipe2000,Fei2020,Ishizuka2022,Raj2024,Brenig2024},
primarily in the context of very different physical questions. We
follow these studies and regularize $\chi_{2}(\omega_{1},\omega_{2})$
along the GEE line by replacing the causal broadening $\eta$ with
a \emph{physical scattering rate} $\Gamma$. The latter can be caused,
e.g., by self-interactions, phonons, or disorder. It remains a free
parameter for the remainder of this study and could, in principle, be also frequency and momentum
dependent. In phenomenologically incorporating such scattering in the
numerical data which we present for $\chi_{2}$, we balance the value
of $\Gamma$ such as to keep the GEE the dominant feature, but to
still highlight all relevant details of the 2D NLRS elsewhere in the
$(\omega_{1},\omega_{2})$-plane.

\begin{figure}[tb]
\centering{}\includegraphics[width=0.99\columnwidth]{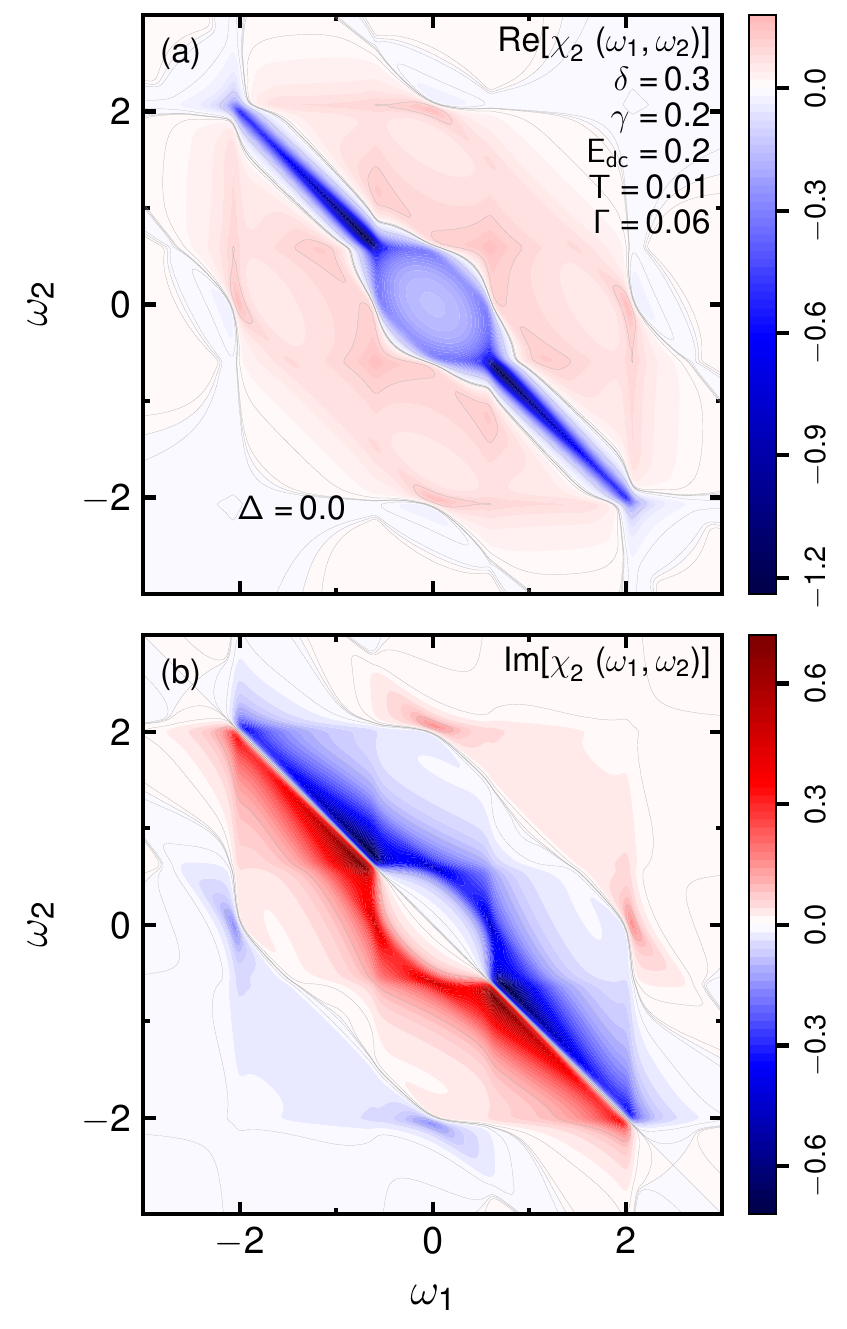}\vskip -.3cm\caption{
Contours of (a) the real and (b) imaginary part of the 2D NRF $\chi_{2}(\omega_{1},\omega_{2})$
in the  XY-limit for a representative set of parameters listed in legend of (a).
Contour lines added at $\pm(.1,.01,.001,0)$ for better visibility of low amplitude
structures. Energies in units of J, as in Eq. (\ref{eq:model}).\label{fig:o2cont}}
\end{figure}

Fig. \ref{fig:o2cont} shows a typical set of contours of $\chi_{2}(\omega_{1},\omega_{2})$
in the 2D frequency plane. The most prominent feature is its antidiagonal
structure which is large regarding its real part and strictly zero
along $\omega_{1}=-\omega_{2}$ regarding its imaginary part. This
reflects the discussion of Eq. (\ref{eq:c2gee2}). The absolute value
of $\mathrm{Re}[\chi_{2}]$ at any point on the line $(\omega,-\omega)$
in the upper panel scales $\propto1/\Gamma$ at small $\Gamma$. It
is straightforward to show that perpendicular, yet close to the line
$\omega_{1}=-\omega_{2}$, the real part of $\chi_{2}$ displays a
Lorentzian shape of width $\Gamma$. This is consistent with the narrow
extent of the region of large $\mathrm{Re}[\chi_{2}]$ perpendicular
to the antidiagonal. See also a similar conclusion for the Kitaev
model, featuring Majorana fermions \citep{Brenig2024}. Therefore,
and as a main result of our work, this implies that scattering rates
of the fractional spinons can be read off from the perpendicular width
of the antidiagonal of the $O(2)$ 2D NRF in the XY-limit. Connecting
to the introduction in Sec. \ref{sec:Introduction},
we mention that the linear response,
i.e., the optical conductivity $\chi_{1}(\omega)$ resulting from
$\tilde{H}^{xxz}$ and $P$, would show a continuum only and reveal
no such information. Finally and quantitatively, for the parameters
used in Fig. \ref{fig:o2cont}, the lower{[}upper{]} two-spinon energy
is at $\min[\max](2\epsilon_{k})\approx0.604[2.04]$. Consistent with
Eq. (\ref{eq:c2gee2}), these two energies are clearly visible, limiting
the regions of large $\mathrm{Re}[\chi_{2}]$ along the antidiagonal.
The smearing of the corresponding energy cut-offs is due to the finite
value of $\Gamma$.

\section{XXZ-case: Renormalized Light-Matter Coupling\label{sec:renlmc}}

In this section we consider spinon interactions. A complete treatment
of $H^{zz}$ requires fully dressing the diagrams of Fig. \ref{fig:diags}
by four-fermion interactions. This is beyond our analytic treatment.
However, a particular subset of interaction processes is accessible.
I.e., when light interacts with the magnetic system, each polarization
operator creates a spinon particle-hole pair. The presence of zz-exchange
turns those pairs into interacting two-spinon states which can be
treated by polarization-vertex corrections adapted to this two-particle
limit. These vertex corrections can also be viewed to generate a
renormalized light-matter coupling. They modify the spinon-pairs'
joint density of states and may induce collective (anti)bound spinon-pair
states, i.e., excitons. This can lead to \emph{qualitative} modifications
of $\chi_{2}$. Our aim is to focus solely on this kind of interaction effect.

Fig. \ref{figRPA} summarizes our corresponding approach formally:
As depicted in Fig. \ref{figRPA}(a), we renormalize the 2D NRF by
vertex corrections. In Fig. \ref{figRPA}(b) these are clarified to
consist of a sum of the bare polarization vertex as well as the renormalizations
of the latter by spinon particle-hole interactions. Finally, Fig.
\ref{figRPA}(c) states that the interactions within this latter particle-hole
pair are treated using a random-phase approximation (RPA). This seems
sufficient for an approach in the two-particle limit which assumes $\Delta$
to be treated perturbatively.

Among the remaining interaction effects not considered, are also the one-spinon
self-energies. Their real part introduces only a \emph{quantitative} and
perturbative change of the spinon dispersion and of $\chi_{2}$.
Their imaginary part introduces a lifetime broadening to $\chi_{2}$
which we subsum phenomenologically into $\Gamma$ as discussed in Sec.
\ref{sec:2DCS-Response-Fun}.

\subsection{Vertex RPA\label{subsec:RPA}}

\begin{figure}[t]
\centering{}\includegraphics[width=0.95\columnwidth]{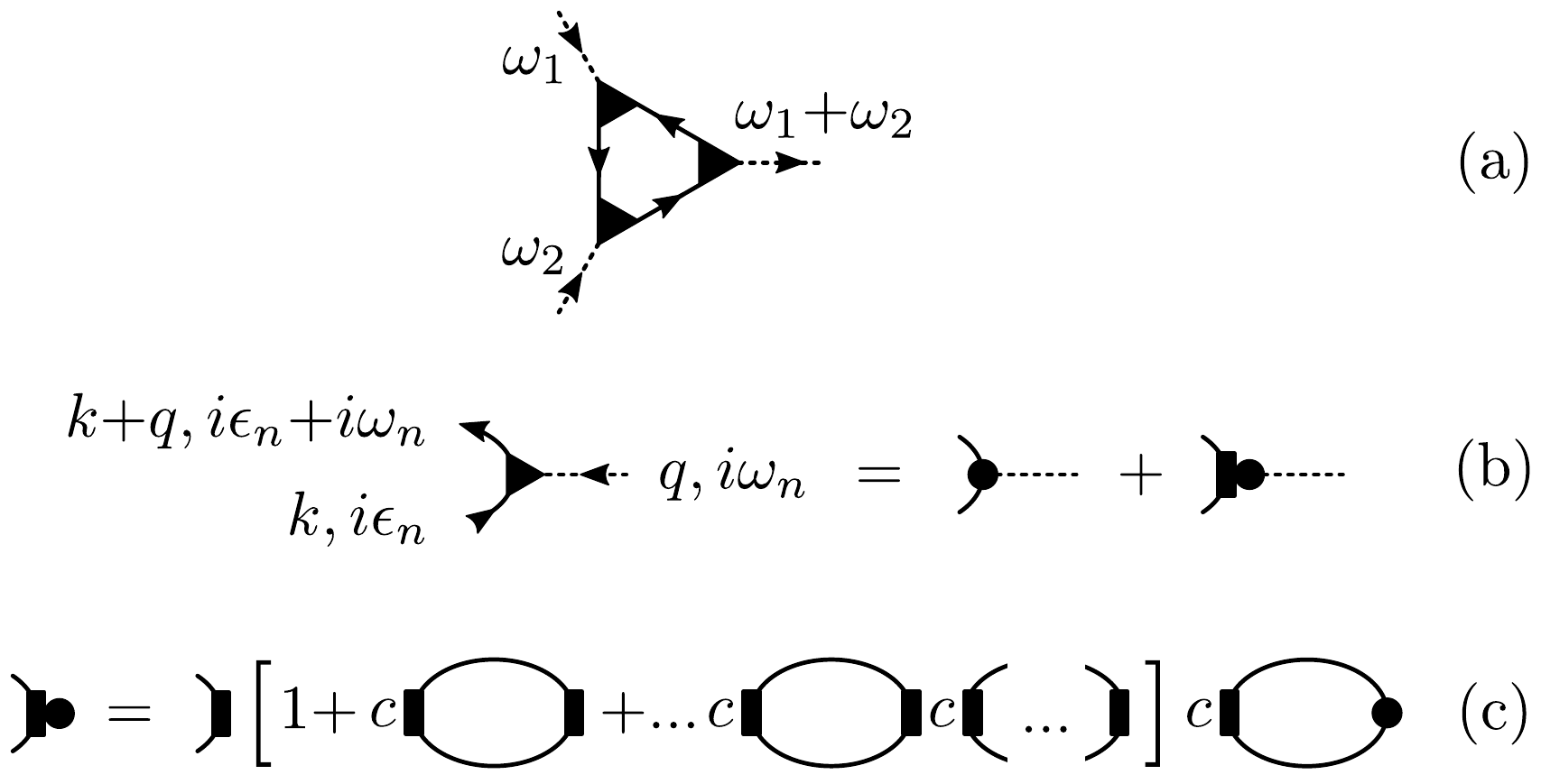}\caption{
Panel (a) Vertex corrected bare diagrams of Fig. \ref{fig:diags}.
Panel (b) The dressed polarization vertex (solid triangle) is a sum
of the bare vertex (solid circle) and the bare vertex coupled to an
interacting spinon particle-hole pair (solid rectangle with circle).
Panel (c) The interacting spinon particle-hole pair is generated by
RPA scattering from the bare four-spinon vertex. This vertex is momentum-separated
into a coupling-constant matrix $c$ and eigenfunctions (solid rectangles,
left and right of $c$). For simplicity, not all arrows are drawn.
In particular, all upper spinon lines contribute a creation (destruction)
operator at their left (right) attachment at any vertex. Similarly
for all lower lines, with creation and destruction interchanged. The
diagrams in panels (b) and (c) refer to the original spinons $a_{\mu,k}^{(\dagger)}$
with $\mu{=}1{,}2$. I.e., the Green's functions $G_{\mu\nu}(k,i\epsilon_{n})$
are non-diagonal with respect to $\mu,\nu$.\label{figRPA}}
\end{figure}

We now detail the treatment of the vertex corrections shown in Fig.
\ref{figRPA}(b) and (c). To organize the RPA of panel (c), we envision
all distinct orderings of the four-fermion interactions in $H^{zz}$
into a product of two particle-hole operators which occur within
an RPA when contracting the operators into particle-hole bubbles
from left to right as in Fig. \ref{figRPA}. There are four such orderings
which we label by ${:}H^{zz}{:}$
\begin{align}
{:}H^{zz}{:}/\Delta= & \sum_{q}v_{q}\,\sum_{k}a_{1,k+q}^{\dagger}a_{1,k}^{\phantom{\dagger}}\,\sum_{k'}a_{2,k'}^{\dagger}a_{2,k'+q}^{\phantom{\dagger}}\label{Vdx}\\
= & \sum_{q}v_{-q}\,\sum_{k}a_{2,k+q}^{\dagger}a_{2,k}^{\phantom{\dagger}}\,\sum_{k'}a_{1,k'}^{\dagger}a_{1,k'+q}^{\phantom{\dagger}}\nonumber 
\end{align}
\vspace{-4mm}
\begin{align*}
\hphantom{{:}H^{zz}{:}/\Delta}= & \sum_{q}(-)v_{k-k'}\,\sum_{k}a_{1,k+q}^{\dagger}a_{2,k}^{\phantom{\dagger}}\,\sum_{k'}a_{2,k'}^{\dagger}a_{1,k'+q}^{\phantom{\dagger}}\\
= & \sum_{q}(-)v_{k-k'}^{\star}\,\sum_{k}a_{2,k+q}^{\dagger}a_{1,k}^{\phantom{\dagger}}\,\sum_{k'}a_{1,k'}^{\dagger}a_{2,k'+q}^{\phantom{\dagger}}\,,
\end{align*}
where $\Delta v_{q}=\Delta(2\allowbreak\cos(q)+\allowbreak2i\delta\sin(q)$
is the Fourier transform of the zz-interaction $\Delta(1+(-)^{l}\delta)\delta_{l,l+1}$
from Eq. (\ref{eq:jwh0}). I.e., there are two direct and two exchange
groupings. Any difference by one-body terms between the groupings
are discarded, as they do not contribute to the contractions into
particle-hole bubbles. To abbreviate the notations, we do not display
global prefactors of $1/N_{k}$, which ensure proper extensivity,
where $N_{k}$ is the number of $k$-points. Such normalizing factors
are assumed to be part of any final $k$-summations. Defining the
operators
\begin{equation}
A_{lm,k,q}^{\dagger}=a_{l,k+q}^{\dagger}a_{m,k}^{\phantom{\dagger}}\,,\label{Almkq}
\end{equation}
Eq. (\ref{Vdx}) can also be written as
\begin{align}
{:}H^{zz}{:}/\Delta= & \sum_{q}v_{q}\,\sum_{k}A_{11,k,q}^{\dagger}\,\sum_{k'}A_{22,k,q}^{\phantom{\dagger}}\label{Adx1}\\
= & \sum_{q}v_{-q}\,\sum_{k}A_{22,k,q}^{\dagger}\,\sum_{k'}A_{11,k,q}^{\phantom{\dagger}}\label{Adx2}\\
= & \sum_{q}(-)v_{k-k'}\,\sum_{k}A_{12,k,q}^{\dagger}\,\sum_{k'}A_{12,k,q}^{\phantom{\dagger}}\label{Adx3}\\
= & \sum_{q}(-)v_{k-k'}^{\star}\,\sum_{k}A_{21,k,q}^{\dagger}\,\sum_{k'}A_{21,k,q}^{\phantom{\dagger}}\,.\label{Adx4}
\end{align}
This notation merely clarifies that all irreducible contributions
for the RPA are of type $\sim\langle\langle A_{x}A_{y}^{\dagger}\rangle\rangle$.

In principle, the RPA is an integral equation. However, it simplifies
enormously if all orderings ${:}H^{zz}{:}$ can be separated with
respect to $k$, $k'$. In that case, the RPA turns into an algebraic
inversion. While the orderings (\ref{Adx1},\ref{Adx2}) are separated
already, those of (\ref{Adx3},\ref{Adx4}) are not. However, the
interaction $v_{k-k'}$ allows for a decomposition in terms of four
eigenfunctions
\begin{align}
-v_{k-k'}= & \nu_{1}h_{1,k}^{\star}h_{1,k'}+\nu_{2}h_{2,k}^{\star}h_{2,k'}\nonumber \\
 & +\nu_{3}h_{3,k}^{\star}h_{3,k'}+\nu_{4}h_{4,k}^{\star}h_{4,k'}\,,
\end{align}
with eigenvalues $\nu_{i}=-2,-2,-1,1$, where
\begin{align}
h_{i,k}^{\star}=( & \cos(k),\sin(k),\delta^{1/2}(\cos(k)+i\sin(k)),\nonumber \\
 & \hphantom{aaa}\delta^{1/2}(\cos(k)-i\sin(k)))\,.
\end{align}
This separates each of the orderings (\ref{Adx3},\ref{Adx4}) into
four distinct channels with respect to $k$, $k'$. Altogether, we
therefore obtain $1{+}\allowbreak1{+}\allowbreak4{+}\allowbreak4=\allowbreak10$
momentum separated orderings of the interaction into products of two
particle-hole operators which have to be contracted mutually to generate
the RPA series. This can be encoded into a vector-matrix notation.
First, we define a vector of 10 particle-hole operators
\begin{equation}
B_{i=1\dots10,q}^{\dagger}=\sum_{k}g_{i,k}^{\star}A_{l(i)m(i),k,q}^{\dagger}\,,\label{Bd}
\end{equation}
where for $i{=}1{,}\dots{,}10$ the indices $l(i)$ and $m(i)$ are
$l(i){=}\allowbreak1{,}\allowbreak2{,}\allowbreak1{,}\allowbreak1{,}\allowbreak1{,}\allowbreak1{,}\allowbreak2{,}\allowbreak2{,}\allowbreak2{,}\allowbreak2$
and $m(i){=}\allowbreak1{,}\allowbreak2{,}\allowbreak2{,}\allowbreak2{,}\allowbreak2{,}\allowbreak2{,}\allowbreak1{,}\allowbreak1{,}\allowbreak1{,}\allowbreak1$
and $g_{i,k}^{\star}=(1,1,\allowbreak h_{1,k}^{\star},\dots,\allowbreak h_{4,k}^{\star},h_{1,k},\allowbreak\dots,\allowbreak h_{4,k})$.
Second, we introduce a $10\times10$ coupling constant matrix
\begin{equation}
\mathbf{c}_{q}=\begin{bmatrix}0 & v_{q}\\
v_{q}^{\star} & 0\\
 &  & c_{3,\dots10}
\end{bmatrix}\,,
\end{equation}
which is nonzero only for the two off-diagonal elements $v_{q}^{(\star)}$
and the diagonal $c_{3,\dots,6}=\allowbreak c_{7,\dots,10}=\allowbreak\nu_{1,\dots,4}$.
With these definitions, we combine all orderings of the interaction
as products of two particle-hole operators into the bilinear form
\begin{equation}
{:}H^{zz}{:}=\Delta\sum_{q}\mathbf{B}_{q}^{\dagger}\mathbf{c}_{q}\mathbf{B}_{q}^{\phantom{\dagger}}\,,\label{irrvtx}
\end{equation}
where $\mathbf{B}_{q}^{\dagger}$ denotes the vector $B_{i=1\dots10,q}^{\dagger}$.
We emphasize once more that the sum on the right-hand side of this
equation is \emph{not} $H^{zz}$. Instead, it comprises all distinct
orderings which arise from $H^{zz}$ in a momentum-separated RPA particle-hole
bubble summation. With this, the expression in the square bracket
$[\dots]$ in Fig. \ref{figRPA}(b) reads
\begin{equation}
\mathbf{D}(i\omega_{n})=(\mathbf{1}-\Delta\mathbf{c}_{q}\langle\langle\mathbf{B}_{q}^{\phantom{\dagger}}\otimes\mathbf{B}_{q}^{\dagger}\rangle\rangle)^{-1}\,,\label{RPA}
\end{equation}
where $\langle\langle\mathbf{B}_{q}\otimes\mathbf{B}_{q}^{\dagger}\rangle\rangle$
denotes the transform to Matsurbara frequencies, $\omega_{n}=2n\pi$T,
of the $10{\times}10$ matrix-bubble $\langle T_{\tau}[B_{i,q}(\tau)\allowbreak B_{j,q}^{\dagger}]\rangle$,
with
\begin{align}
\lefteqn{\langle T_{\tau}[B_{i,q}(\tau)B_{j,q}^{\dagger}]\rangle=}\nonumber \\
 & \hphantom{{aaa\}}}\sum_{k,k'}g_{i,k}g_{j,k}^{\star}\langle T_{\tau}[A_{l(i)m(i),k,q}^{\phantom{\dagger}}(\tau)A_{l(j)m(j),k',q}^{\dagger}]\rangle\,.\label{AAbub}
\end{align}
In diagonal particle representation of the $H_{0}$, this is
\begin{align}
\lefteqn{\langle T_{\tau}[B_{i,q}(\tau)B_{j,q}^{\dagger}]\rangle=}\nonumber \\
 & \sum_{k,\mu\nu\lambda\sigma}\left[g_{i,k}g_{j,k}^{\star}U_{m(i)\mu}^{\star}(k)U_{l(i)\nu}^{\phantom{\star}}(k{+}q)U_{l(j)\lambda}^{\star}(k{+}q)\times\right.\nonumber \\
 & \hphantom{\sum_{k}a}\left.U_{m(j)\sigma}^{\phantom{\star}}(k)\,\langle T_{\tau}[d_{\mu,k}^{\dagger}(\tau)d_{\nu,k{+}q}^{\phantom{\dagger}}(\tau)d_{\lambda,k{+}q}^{\dagger}d_{\sigma,k}^{\phantom{\dagger}}]\rangle\right],\label{digbub}
\end{align}
where momentum conservation has been applied. Next, we transform
to Matsurbara frequencies. In doing so, we use that (i) only finite
input frequencies are relevant and (ii) $q{\rightarrow}0$ applies.
With the latter two points, intraband bubbles can be discarded, and we obtain
\begin{align}
\lefteqn{\langle\langle\mathbf{B}_{0}\otimes\mathbf{B}_{0}^{\dagger}\rangle\rangle_{ij}=\int_{0}^{\beta}e^{i\omega_{n}\tau}\langle T_{\tau}[B_{i,q}(\tau)B_{j,q}^{\dagger}]\rangle d\tau|_{q\rightarrow0}=}\nonumber \\
 & \hphantom{aaaaaaaaa}\sum_{k,\mu}\left[g_{i,k}g_{j,k}^{\star}\,U_{m(i)\mu}^{\star}(k)U_{m(j)\mu}^{\phantom{\star}}(k)\,\times\right.\nonumber \\
 & \hphantom{aaaaaaaaa}\hphantom{\sum_{k}a}\left.U_{l(i)\bar{\mu}}^{\phantom{\star}}(k)U_{l(j)\bar{\mu}}^{\star}(k)\,\frac{2f(\epsilon_{\mu,k})-1}{i\omega_{n}+2\epsilon_{\mu,k}}\right]\,,\label{digbubW}
\end{align}
where we reuse the notation $\epsilon_{\mu,k}$, from Eq. (\ref{eq:xydig})
and define $\bar{\mu}=\bar{1}(\bar{2})=2(1)$. Next, we turn to the
rightmost initial bubble in Fig. \ref{figRPA}(b). This is a rank
10$\times$1 tensor comprising the 10 momentum-separated particle-hole
channels, $\mathbf{B}_{q}$, in the first dimension and a sum over
all bare spinon-photon vertices, $P^{\dagger}$, in the second
\begin{align}
\!\! & \mathbf{BP}(i\omega_{n})_{i}\equiv\langle\langle\mathbf{B}_{0}\otimes P^{\dagger}\rangle\rangle_{i}=\nonumber \\
\!\! & \hphantom{aaa}\sum_{k,\mu}g_{i,k}U_{m(i)\mu}^{\star}(k)U_{l(i)\bar{\mu}}^{\phantom{\star}}(k)\,p_{\bar{\mu}\mu}^{\star}(k)\frac{2f(\epsilon_{\mu,k})-1}{i\omega_{n}+2\epsilon_{\mu,k}}\,.\label{startbub}
\end{align}
Again, only interband bubbles contribute, and $q=0$. Processing Eqs.
(\ref{digbubW}) and (\ref{startbub}) further by using that $p_{\bar{\mu}\mu}(k)=-p_{\bar{\mu\mu}}(k)$
and $\epsilon_{\mu,k}=-\epsilon_{\bar{\mu},k}$ does not lead to relevant
simplifications.

Finally, the leftmost particle-hole vertex of Fig. \ref{figRPA}(b)
remains. This is simply a rank 1$\times$10 tensor of all channels
of particle-hole operators, i.e. Eq. (\ref{Bd}), which contracts
the left end of square bracket $[\dots]$ in Fig. \ref{figRPA}(b)
back to a sum of operators. In the diagonal particle representation
it reads
\begin{align}
B_{i,q=0}^{\dagger} & =\sum_{k,\mu\nu}g_{i,k}^{\star}U_{l(i)\mu}^{\star}(k)U_{m(i)\nu}^{\phantom{\star}}(k)d_{\mu,k}^{\dagger}d_{\nu,k}^{\phantom{\dagger}}\nonumber \\
 & \equiv\sum_{\mu\nu}R_{\mu\nu,i}d_{\mu,k}^{\dagger}d_{\nu,k}^{\phantom{\dagger}}\,,\label{Pr}
\end{align}
where, to formulate Fig. \ref{figRPA}(b) analogous to the bare spinon-photon
vertex which has rank 2$\times$2 in terms of $\mu\nu$, we have
defined the 2$\times$2$\times$10 tensor $R_{\mu\nu,i}$. Altogether,
we obtain
\begin{equation}
\smash{{\begin{array}{c}
\\\includegraphics[scale=.32]{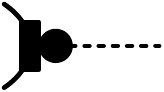}\\
\\\end{array}}}=\Delta\sum_{\mu\nu}\mathbf{R_{\mu\nu}}\cdot\mathbf{D}(i\omega_{n})\cdot\mathbf{c}_{0}\cdot\mathbf{BP}(i\omega_{n})\,d_{\mu,k}^{\dagger}d_{\nu,k}^{\phantom{\dagger}}\label{renvtx}
\end{equation}
for the contribution to the polarization operator which is dressed
by spinon pair interactions, i.e., the rightmost vertex in Fig. \ref{figRPA}(b).
This concludes the derivation of the expression for the vertex correction.

\begin{figure}[tb]
\centering{}\includegraphics[width=0.82\columnwidth]{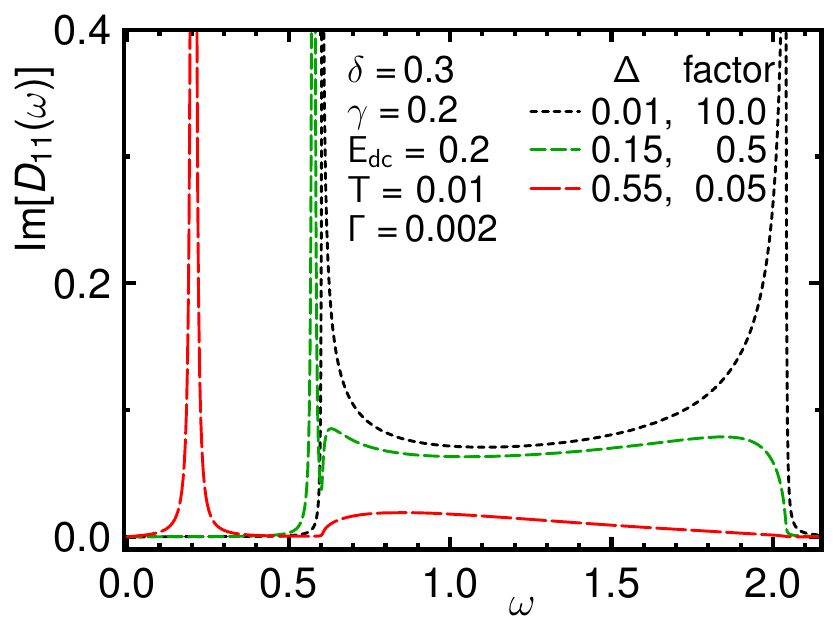}$\hphantom{aaa}$\vskip -.3cm\caption{
Spectrum of the $11$ matrix-element of the RPA $10\times10$ series
$\mathbf{D}(\omega)$ for a representative set of parameters, varying $\Delta$,
as listed in legend. $y$-axis values are scaled by arbitrary factors listed. \label{fig:BndSt}}
\end{figure}

\subsection{Discussion\label{subsec:xxzdisc}}

Now we discuss the impact of the vertex renormalization on the 2D
NRF. To this end, we exemplify the frequency dependence which is
introduced by the RPA into the dressed polarization vertices Fig.
\ref{figRPA}(b). This can be seen best by focusing on the spectrum
of one of the $10\times10$ matrix elements of $\mathbf{D}(\omega)\equiv\mathbf{D}(i\omega_{n}\rightarrow\omega+i\Gamma)$,
e.g., $\mathrm{Im}[D_{11}(\omega)]$, which we show in Fig. \ref{fig:BndSt}.
In this figure we focus on a very small value of the imaginary broadening
such as to display all formal properties of the spectrum. For the
remaining parameters of the spin system, we choose the same set as
in Fig. \ref{fig:o2cont}, however, our conclusions are independent
of this choice.

\begin{figure}[tb]
\centering{}\includegraphics[width=0.99\columnwidth]{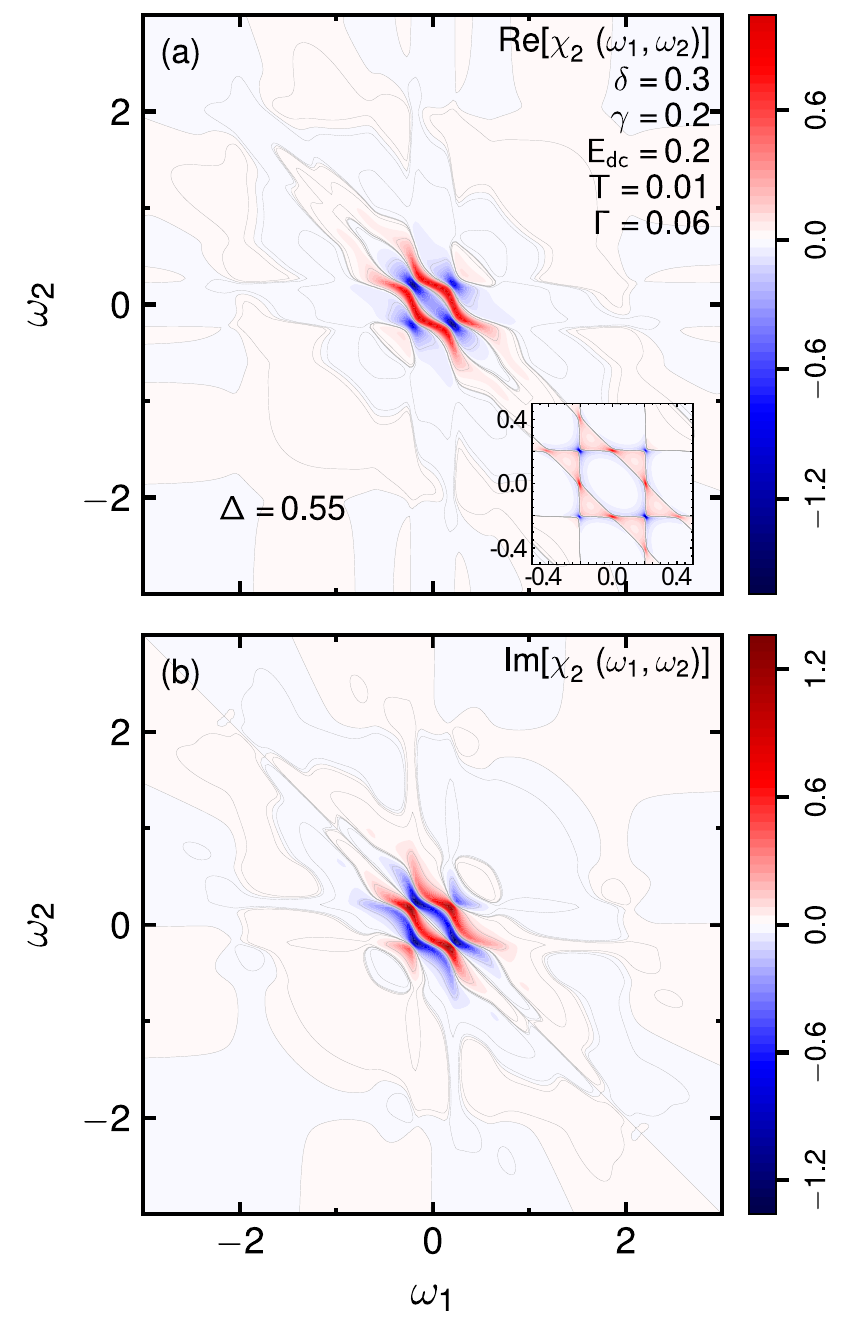}\vskip -.3cm\caption{
Contours of (a) the real and (b) imaginary part of the 2D NRF $\chi_{2}(\omega_{1},\omega_{2})$
for the XXZ-case for a representative set of parameters listed in legend of (a).
Contour lines added at $\pm(.1,.01,.001,0)$ for better visibility of low amplitude
structures. Energies in units of J, as in Eq. (\ref{eq:model}).
Inset in (a): Central region of the NRF with $-0.5\protect\leq\omega_{1,2}\protect\leq0.5$
for a reduced broadening of $\Gamma=0.01$. \label{fig:o2combo}}
\end{figure}

This figure demonstrates a main point of this work, namely that for sufficiently
large coupling constants $\Delta$ of $H^{zz}$, the vertex acquires a strong and
qualitatively relevant mid-gap structure.  We first note that each plot of
$\mathrm{Im}[D_{11}(\omega)]$ in Fig. \ref{fig:BndSt} has been scaled by some
arbitrary factor, as listed, in order to illustrate all features on the same
$y$-axis.  For very small $\Delta=0.01$, the figure shows that the spectral weight of
the RPA contribution is also very small - as to be expected - and essentially
displays van Hove singularities at the zero-momentum two-spinon band edges, i.e.,
$\approx0.604[2.04]$, noted at the end of
Sec. \ref{sec:2DCS-Response-Fun}. Increasing $\Delta$ to $0.15$, a sharp, bound-state
structure is barely starting to split off from the two-spinon continuum and the van
Hove singularities change from a divergent nature to a decreasing cusp behavior. We
strongly caution that fine-tuning of the imaginary broadening $\Gamma$ and the
integration mesh allows for the resolution of a bound state very close to the bottom
of the two-spinon continuum, even for much smaller values of $\Delta$. This suggests
that the critical value for its occurrence is actually $\Delta=0$ for $\Gamma=0$ in
the thermodynamic limit. We do not dwell upon this detail here.
Finally, for the largest of the three $\Delta$ depicted,
i.e. $\Delta=0.55$, which is still reasonably less that one, the
spectrum changes drastically. Namely, the two-spinon bound state is
now a collective excitation well separated from the continuum and
deep within the dimerization-induced spin gap. For completeness, we
mention that, as usual, $\mathrm{Im}[D_{11}(-\omega)]=-\mathrm{Im}[D_{11}(\omega)]$.
The existence of the bound states is fully consistent with past investigations
of dimerized spin chains, where similar results have been reported
in field theoretical work \citep{Uhrig1996}, as well as in exact
diagonalization studies \citep{Fledderjohann1997,Augier1997}. For
the remainder of this work, we will focus on parameters in which the
bound state is well separated from the two-spinon continuum.

Now we contrast the 2D NRFs with and without vertex corrections against
each other. To this end, we remain with the parameters of Fig. \ref{fig:o2cont},
and for the case with vertex corrections, we set $\Delta=0.55$ as
in Fig. \ref{fig:BndSt}. For this value of $\Delta$, contours of
the vertex corrected $\chi_{2}(\omega_{1},\omega_{2})$ are depicted
in Fig. \ref{fig:o2combo} on the same $\omega_{1},\omega_{2}$-scale
as in Fig. \ref{fig:o2cont}. Remarkably, and as another main point
of this work, the extent of the dominant region of $\chi_{2}$ has
strongly contracted within the 2D plane for the interacting case.
This can be understood in terms of the RPA bound state. From the data
of Fig. \ref{fig:BndSt}, an energy of the bound state of $\epsilon_{B}\approx0.205$
can be read off. In terms of the incoming and outgoing frequencies
of the diagram in Fig. \ref{figRPA}(a), i.e., $\omega_{1,2}$ and
$\omega_{1}+\omega_{2}$, this implies six lines of resonance in the
2D plane, namely, $\omega_{1}=\pm\epsilon_{B}$, $\omega_{2}=\pm\epsilon_{B}$,
and $\omega_{1}+\omega_{2}=\pm\epsilon_{B}$. These have twelve crossing
points that are related to bound-states simultaneously excited within
two of the vertices, i.e., a particularly strong response. Obviously,
the area spanned by this response is strongly contracted with respect
to the original energy scale $J\equiv1$, and is confined to within
the gap. To explicitly display this, Fig. \ref{fig:o2combo}(a) contains
an inset of the low-frequency region of the same $\mathrm{Re}\chi_{2}(\omega_{1},\omega_{2})$,
however, for a very small imaginary broadening $\Gamma$. This inset
clearly shows the anticipated twelve fingerprints of the collective
state from the vertex corrections. In practice however, and as for
the GEE-line, intrinsic and extrinsic damping of various kinds are
very likely to broaden such very sharp features and rather merge them
into smoother contours such as those depicted in panels (a), (b) of
the figure.

Finally, we mention the role of temperature. Since $T$ enters all of our
calculations only via Fermi-blocking factors $2 f_k {-}1$ its impact is weak and
only sets in at temperatures significantly higher than the dimerization
gap. Therefore, we have not analyzed the $T$-dependence.  \vspace{-3mm}

\section{Summary\label{sec:sum}}

In conclusion, we showed that 2DCS in magnets results not only from
a coupling of magnetic fields to the magnetic moments, but equally
well from effective electric dipole moments related to the MEE which
couple to external electric fields. For that purpose, we considered
the $O(2)$ 2D NRFs for the XXZ-chain containing effective dipoles
of the KNB type. For the noninteracting XY-case, we found that the
2D NRFs do not only display multiparticle continua similar to linear
spectroscopies, but, most important, can provide information about
the individual elementary excitations lifetimes. For fractionalized
systems, the latter are inaccessible to linear response probes. While
at $O(2)$ information about the elementary excitations is linked
to the GEE line of the 2D NRF, we suspect that at higher orders of
response similar lines exist at other locations in frequency space.
For the XXZ case, we analyzed the interaction effects resulting from
multiple scattering of the spinon pairs created by the incoming light.
While this is certainly only a subset of possible renormalizations
due to $H^{zz}$, we have shown that collective bound spinon-pairs
states arising from this can lead to large quantitative and qualitative
changes of the 2D NRFs. Unbiased numerical analysis of such phenomena
would potentially be useful.

\begin{acknowledgments}
Anna Keselman and Natalia Perkins contributed very valuably to the early stages of
this research. Fruitful discussions with Ciaran Hickey, Johannes Knolle and Peter
Orth are gratefully acknowledged. Research of W.B. was supported in part by the DFG
through Project A02 of SFB 1143 (project-id 247310070) and by NSF grant PHY-2210452
to the Aspen Center for Physics (ACP).  W.B. acknowledges kind hospitality of the
PSM, Dresden.
\end{acknowledgments}

\end{document}